\documentclass[sn-mathphys]{sn-jnl}

\makeatletter
  \@addtoreset{chapter}{part}
  \@addtoreset{@ppsaveapp}{part}
\makeatother

\jyear{2021}%

\usepackage{amsmath,amssymb,amstext,amsthm} 
\usepackage{verbatim,graphicx,pdfpages}
\usepackage{mathtools}
\usepackage{soul}
\usepackage[multiple]{footmisc}
\usepackage{algorithm}
\usepackage{algcompatible}
\usepackage{setspace}
\usepackage{euscript}
\usepackage{enumerate}
\usepackage{color}
\usepackage{enumitem}
\setlist{leftmargin=5.5mm}
\usepackage{adjustbox}
\usepackage{dashbox}
\usepackage{lineno}
\usepackage{longtable}
\usepackage{lscape}
\usepackage{verbatim}
\usepackage{xspace} 
 
\usepackage{color} 
\usepackage{eurosym}

\newtheorem{definition}{Definition}

\newcommand{\bb}{\textbf}
\newcommand{\bbb}{\boldsymbol}
\newcommand{\tn}{\textnormal}
\newcommand{\rz}{{\mathbb{R}}}

\begin{document}
\pagestyle{plain}

\title[Integrating multiple sources of ordinal information]{Integrating multiple sources of ordinal information in portfolio optimization}


\author[1]{\fnm{Eranda} \sur{\c{C}ela}}\email{cela@math.tugraz.at}
\author[2]{\fnm{Stephan} \sur{Hafner}}\email{}
\author[3]{\fnm{Roland} \sur{Mestel}}\email{roland.mestel@uni-graz.at}
\author*[2]{\fnm{Ulrich} \sur{Pferschy}}\email{ulrich.pferschy@uni-graz.at}

\affil[1]{\orgdiv{Department of Discrete Mathematics}, \orgname{TU Graz}, \orgaddress{\street{Steyrergasse 30}, \city{Graz}, \postcode{8010}, \country{Austria}}}
\affil*[2]{\orgdiv{Department of Operations and Information Systems}, \orgname{University of Graz}, \orgaddress{\street{Universitaetsstrasse 15}, \city{Graz}, \postcode{8010}, \country{Austria}}}
\affil[3]{\orgdiv{Department of Banking and Finance}, \orgname{University of Graz}, \orgaddress{\street{Universitaetsstrasse 15}, \city{Graz}, \postcode{8010}, \country{Austria}}}

\abstract{ 
Active portfolio management tries to incorporate any source of meaningful information into the asset selection process.
In this contribution we consider multiple qualitative views specified as total orders of the expected asset returns and discuss two different approaches for incorporating this input in a mean-variance portfolio optimization model. 
In the robust optimization approach we first compute  a posterior expectation of asset returns for every given total order by an extension of the Black-Litterman (BL) framework.
Then these expected asset returns are considered as possible input scenarios for robust optimization variants of the  mean-variance portfolio  model (max-min robustness, min regret robustness and soft robustness). 
In the order aggregation approach 
rules from social choice theory (Borda,  Footrule, Copeland, Best-of-k and MC4) are used to aggregate the total order in a single ``consensus total order''.
Then expected asset returns are computed for this ``consensus total  order'' by the extended BL framework mentioned above. 
Finally, these expectations are used as an input of the classical mean-variance optimization. 
Using data from EUROSTOXX 50 and S\&P 100 we empirically compare the success of the two approaches in the context of portfolio performance analysis and  observe that in general aggregating orders by social choice methods outperforms robust optimization based methods for both data sets.
}


\maketitle

\section{Introduction}
\label{sec:intro}

\modulolinenumbers[5]

{This paper addresses the question of how to incorporate qualitative information about future stock prices into portfolio optimization. 
We apply a recent approach by \cite{CeHaMePfe2021JBF} that extends the Black-Litterman model (\cite{black1991asset,black1992global}; hereafter BL) by allowing views that simply rank assets or asset classes according to their expected future performance.\footnote{Note that a view is generally defined as an uncertain statement about the expected return of one or more securities.} However, while \cite{CeHaMePfe2021JBF} consider a single source of ordinal information (e.g., the beliefs of a single financial expert or a ranking based on a single technical indicator of stock prices), we introduce a significant extension by allowing multiple ordering relations to input diverging views about different assets.
We pursue two different approaches for dealing with multiple orderings: On the one hand, we use different models from robust optimization considering each ordering as a possible scenario. 
On the other hand, we apply methods from social choice theory to aggregate multiple rankings into a single consensus ranking.
We will show how both approaches can be used to integrate multiple rankings into mean-variance (MV) analysis. To the best of our knowledge, this is the first time that aggregation methods from social choice theory have been integrated into the world of optimal asset allocation. In addition to formal explanations, we compare the performance of these approaches in a simplified setting using historical stock market data. This allows us to formulate recommendations on which method is preferable when considering multiple pieces of qualitative information in MV portfolio optimization.} 


\subsection{Mean-variance portfolio optimization and the Black-Litterman approach}
\label{sec:MV_BL}

The relevance of incorporating multiple exogenous insights, such as portfolio managers' judgments, into formal models of portfolio management was already highlighted in Harry Markowitz's seminal work on portfolio selection, commonly referred to as mean-variance (MV) optimization or Modern Portfolio Theory (MPT):

\begin{quote}
\textit{''Various types of information concerning securities can be used as the raw material of a portfolio analysis. One source of information is the past performance of individual securities. A second source of information is the beliefs of one or more security analysts concerning future performances."}(\cite{Markowitz1959}, p.~3)
\end{quote}


MPT (\cite{Markowitz_1952}) is undoubtedly the most important quantitative framework for portfolio optimization. 
While the theory is intuitive and coherent, reputation and implementation have suffered among practitioners due to numerous difficulties.  
In particular, to construct alternative portfolios with different combinations of risk and return, with one portfolio being the optimal (or \textit{efficient}) one, investors must provide estimates of  expected returns, volatilities and correlations for all securities in the investment universe under consideration. However,
early theory remains vague on how to specifically estimate these inputs to the investment process. Traditionally, the use of historical market data has been suggested. Unfortunately, using historical distributions to estimate expected returns and the covariance matrix is prone to measurement errors, which can lead to significant discrepancies between \textit{ex-ante} and \textit{ex-post} optimal portfolio weights for a variety of reasons. Empirical evidence of the poor out-of-sample performance of the classical MV optimization model is provided, among others, by \cite{Michaud}, \cite{Best}, 
and \cite{chopra1993}. 

To mitigate the adverse effects of neglecting uncertainty in the inputs and solving the portfolio optimization problem in the original MV framework as a deterministic problem, several approaches have been taken  in the literature  (for a concise overview, we refer to \cite{Kolmetal2014}). 
Some of these approaches focus on modifying the constraints and/or the goal objective function, e.g.\ by imposing restrictions on portfolio weights (e.g., \cite{Michaud}, \cite{Levyetal2014}) and incorporating higher moments and tail-risk measures (e.g., \cite{Harveyetal2010}, \cite{Lassanceetal2021}).
Other approaches target appropriate modifications to the input of the portfolio selection problem including robust optimization methods (e.g., 
\cite{DeMiguel2009}, \cite{Huangetal2010}, 
Bayesian statistics (e.g., \cite{Pastoretal2000}, \cite{Bodnaretal2022}) and the Black-Litterman model (\cite{black1991asset,black1992global}. 
The latter provides the framework for our analysis.    

  

The original BL-approach assumes that the ``true'' expected asset returns $\bbb\mu$  are both,  unknown and \textit{random}. 
Allowing for this uncertainty in estimation, the model starts with equilibrium risk premia as the neutral reference point (the \textit{prior distribution}) for expected returns, derived  from  an asset pricing model such as the CAPM using reverse optimization. 
The prior can be interpreted as the \textit{market view} on the portfolio optimization process, since all investors who do not have specific views on expected asset returns will hold the same optimal portfolio, the market portfolio. 
In a second step, \cite{black1991asset} assume that investors formulate their own individual views on (at least some of) the securities in the investment universe.
These can be views on absolute returns of individual assets (e.g.\ ``the expected one-year return on security X is 3.5\%'') or relative predictions about the difference in returns between two or more assets (e.g.\ ``security X will outperform security Y by 1\% over the next 6 months''). 
As with the prior distribution for $\bbb\mu$, investor individual views are assumed to be stochastic, with uncertainty inversely proportional to an investor's confidence in his/her views.

After separately specifying the prior distribution of expected returns implied by the market equilibrium and an investor's views on expected asset returns, the BL-model applies Bayes' theorem to combine these two sources of information to derive the posterior distribution of $\bbb\mu$. 
The lower an investor's confidence in his/her views, the closer his/her posterior distribution of expected returns will be to that implied by the market equilibrium. 
On the other hand, the more (s)he trusts her own views the more expected returns will tilt away from market equilibrium in the direction of the views.   
The vector of expected returns derived from the BL-model ($\bbb\mu_{\tn{BL}}$) is obtained as the conditional expectation of the prior distribution given the investor's views. 
These estimates finally  enter the MV portfolio optimization process.

\subsection{Qualitative views}
\label{sec:qual_views}

With respect to investor's views, the traditional BL-model makes two assumptions. 
First, views, whether formulated in absolute or relative terms (see above), are assumed to be metrically scaled variables (we will refer to such views as \textit{quantitative} views hereafter). Although this assumption simplifies the calculation of $\bbb\mu_{\tn{BL}}$, it is unlikely that this kind of views will prevail in practice.
Portfolio managers and financial analysts are likely to feel more comfortable expressing \textit{qualitative} views, e.g., by simply ranking their expectations concerning the future performance of individual stocks, different asset classes, several sectors of the economy, or individual or multiple stock markets in different geographic regions. 
In this sense \cite{Fabozzi2007}, p.~233, state that ``...trading strategies ... rather just provide \textit{relative} rankings of securities that are predicted to outperform/underperform other securities.'' 
Further (p.~234):``Clearly, it is not an easy task to translate any of these relative views into the inputs required for the modern portfolio theoretical framework.''

Second, the BL-model and most of its extensions assume that it is individual views being processed as input. 
This is in contrast to team-based portfolio management, which has become very popular in recent years, especially among institutional investors such as mutual funds (e.g., \cite{Patel2017}). 
In a group such as an investment committee, group members usually have different opinions and perspectives on the possible future performance of individual assets, sectors, or even entire markets. 
In such a situation, the team's decisions inevitably will be compromises between these different opinions (e.g., \cite{Sah1986}). 
Multiple views may also matter to individual investors. 
Consider an investor who incorporates the views of several financial analysts into his/her portfolio selection process in addition to his/her own views. 
In both cases, the assumption that only individual views are processed to derive the posterior distribution of expected asset returns is not appropriate. 
However, considering multiple views immediately raises the question of how to combine these views to derive an optimal portfolio. 

\subsection{Contribution}
\label{sec:contribution}

We contribute to the literature dealing with the integration of  qualitative views into portfolio optimization using the parametric BL-framework. 
Following the argument of \cite{Fabozzi2007} above, we apply an approach recently proposed by \cite{CeHaMePfe2021JBF} that translates qualitative views, expressed as  ordering relations between asset returns, into quantitative estimates of expected returns. 
Moreover, we extend \cite{CeHaMePfe2021JBF} by allowing multiple qualitative views and apply  
two different approaches to incorporate these views into portfolio optimization: the first one (I)  is based on robust optimization and the second one (II) on social choice theory. 
We are not aware of any work that has applied methods from social choice to portfolio optimization before.\footnote{It should be noted that \cite{simonian14} contains a first pointer into the direction of social choice theory. The paper addresses a  specific situation in which each view includes a set of stochastically related components.  The goal is to aggregate the views  such  that the components of the aggregated view fulfill the stochastic relation to the greatest extent possible.  The components of the aggregated view are seen in analogy  to an  allocation vector in cooperative game theory and a Shapley-value-based heuristic  is proposed to  compute them.}

In (I), we consider $K$ different ranking orders (e.g., views of $K$ financial analysts) and apply the approach discussed in \cite{CeHaMePfe2021JBF} to compute a vector  of estimated returns for each of them. 
These $K$ (different) vectors of expected returns are considered as  the possible scenarios of expected returns in the MV portfolio optimization model (MVO). 
We solve three robust variants of the MVO which are obtained by applying the  following concepts of discrete robustness: \textit{max-min robustness}, \textit{min-max regret robustness}, and \textit{soft robustness}. 
Depending on the concept of discrete robustness involved, we obtain different   robust optimization-based methods as representatives of this ``first estimate, then aggregate'' approach.

In (II) we follow a completely  different approach. 
Given $K$ views, we first aggregate them into a single ``consensus'' ordering by  applying  some  method from social choice theory (varying among the \textit{Borda rule}, the \textit{Footrule aggregation}, the \textit{Best-of-k-Algorithm}, the \textit{Copeland method} and the \textit{MC4-Algorithm}). 
We then apply the approach discussed in \cite{CeHaMePfe2021JBF} to  compute a  vector of estimated returns  from this single total order. 
Finally, we use this   vector as input to the classical MVO and  compute the respective optimal portfolio. In summary,  in (II), we first aggregate the orderings and then generate the expected return estimator. Depending on the aggregation rule from social choice theory we obtain different social choice-based methods  as representatives of  this ``first aggregate, then estimate'' approach. 

The main advantage of our proposed aggregation methods from  (I) and (II)  is that they are purely deterministic, in the sense that they  do not use probabilistic assumptions in the aggregation process itself.
This is in contrast to well known approaches used to aggregate multiple  views in portfolio optimization, such as the classical Black-Litterman approach~(BL) in \cite{black1992global}, the copula-opinion pooling technique in \cite{Meucci06}, or the generalized Black-Litterman model in \cite{Chen2022}\footnote{\label{footnote_Chen}\cite{Chen2022} consider multiple quantitative  views and generalize the classical  BL approach to account for biases in both the prior distribution and the expert's views. The biased views and the biased prior are aggregated into a posterior distribution by using historical forecast data in a Bayesian hierarchical model.}, all of which consider quantitative views, and the entropy pooling approach in  \cite{Meucci08flex}, which also treats  qualitative views.
All these approaches (implicitly) assume a stochastic dependence structure among the different  views. 
Essentially, this is required because in these approaches  aggregation is an integral part of  the estimation process  and the latter is based on distributional assumptions. The proposed approaches (I) and (II) decouple the aggregation from the estimation process, so that no assumptions about the stochastic dependencies between the different views need to be made.  
This is an obvious advantage since it is not necessary to figure out suitable distributional assumptions and then calibrate the distribution's parameters.    

Another approach to generating the inputs of the MVO (or a more general portfolio optimization model) by ``blending''  equilibrium arguments with (more general) investor's views, without resorting to specific distributional assumptions,  is discussed in \cite{Bertsimas2012}. 
This  is a fairly general inverse optimization approach that can incorporate any view that can be expressed as a linear matrix inequality. 
Thus, it can incorporate views addressing  risk measures or risk factors in addition to views addressing  expected returns.  
This approach  is  conceptually and methodologically quite different from the other approaches mentioned in this paper and we refer to \cite{Bertsimas2012} for further details.

{Finally, to determine which of our different approaches from robust optimization (I) and social choice theory (II) gives the best result in portfolio optimization, we compare the performance of all approaches in a MV framework. 
To this end, we consider two performance measures of the in-sample optimal portfolios, the Sharpe Ratio and the Certainty-Equivalent return, using historical financial market data.
From a practical point of view, establishing  appropriate ordering relations between single assets or asset classes is undoubtedly challenging. 
However, this task should not necessarily be expected to be performed by experts relying on their personal views, but could be solved, for example, using fundamental analysis or technical analysis, two methods commonly used in practice to sort and pick stocks. 
It should be noted that the objective of this paper is not to show different ways of deriving such rankings, but to illustrate different approaches suitable for processing multiple rankings (obtained from any conceivable source) and compare the performance of these approaches in a MV portfolio framework. 
For this reason, we use synthetic views which are rankings of expected asset returns generated uniformly at random at some distance \textit{d} from the realized ordering (which is naturally known in an in-sample setting). This distance value can be considered as a control parameter for the quality of the information represented in these views. The second parameter in our computational experiments is the confidence level in the views, denoted by \textit{c}. We evaluate the performance of the approaches (I) and (II) for different combinations of values of the parameters \textit{c} and \textit{d}. In a first step we compare the methods within groups (I) and (II) separately, and in a second step we compare the best performing methods from each group with each other.}


Our findings can be summarized as follows.
In general, the best methods of group (I) are the min-max regret robustness and the soft robustness. 
In group (II), the methods based on the Borda rule and on the Footrule aggregation clearly perform better than the others. The comparison across the two groups of methods (I) and (II) reveals that the best social choice-based methods outperform the best robust optimization-based methods for almost all parameter settings, albeit the absolute difference between the performance measures is not very accentuated. Thus, our results show that an investor considering incorporating multiple qualitative views into the portfolio allocation decision should use a method from social choice and aggregate first, and then estimate, rather than the other way around.

\bigskip
The remainder of the paper is organized as follows. 
Section \ref{sec:formal} prepares the formal framework for our analysis.
Section \ref{sec:robust} presents different methods from robust optimization for deriving an optimal portfolio given multiple expert views.
Section \ref{sec:social} relates to selected methods from social choice theory that can be applied in a first step to compute a single consensus total order of expected asset returns from different orders. 
In a second step the aggregated order feeds into the portfolio selection process. 
Section \ref{sec:comp} presents the results of our computational experiments based on historical stock market data. 
We conclude and discuss future work in section \ref{sec:conclusion}.

\section{Formal framework for ordinal information}
\label{sec:formal}

In this section we introduce the formal framework and briefly review the
extension of the BL model derived in \cite{CeHaMePfe2021JBF}.
We consider a universe of assets $1,\ldots,n$ from which an investor has to  choose a portfolio represented by a weight vector
$\bb w = (w_1,\ldots,w_n)^\intercal$ with $\sum_{i=1}^n w_i = 1$,
where $w_i$ represents the fraction of capital invested in asset $i$. 
We assume that $ w_i \geq 0\ (i=1,\ldots,n)$, i.e.\ no short selling is allowed.
Let $\bbb \mu = (\mu_1,\ldots,\mu_n)^\intercal = \tn E(\bb R)$ be the expected value of the random vector of asset returns $\bb R = (R_1,\ldots,R_n)^\intercal$ with an $n \times n$ 
covariance matrix $\bbb\Sigma =\tn{Cov}(\bb R)$.
We assume that $\bbb\mu$ is a normally distributed random vector
\begin{equation}
\bbb\mu \sim \tn N_n(\bbb\pi, \tau\bbb\Sigma)
\end{equation}
with mean vector $\bbb\pi$ and covariance matrix $\tau\bbb\Sigma $. $\bbb\Sigma$ is estimated from historical data and  $\bbb\pi$ is determined by invoking  CAPM equilibrium arguments as described for example in \cite{CeHaMePfe2021JBF}. The scale parameter $\tau$ indicates  the uncertainty of the prior,  the smaller the value of $\tau$, the  higher the investor's confidence in the estimation of the implied equilibrium return $\bbb\pi$. 

The classical BL model considers  quantitative views which express statements about the expected return of  linear combinations  of the considered assets.\footnote{For a more detailed review of the BL model we refer to \cite{meucci2010black} or \cite{walters2014black}.} These statements   are represented as equalities of the  form $\bb P\bbb \mu =\bb v$, where $\bb P$ is  a so-called pick matrix and   $\bb v$ is a declared outcome.
For example, the  view that asset 1 will outperform asset 3 by $2\%$ and asset 2 will outperform asset 4 by $1\%$ can be modeled by setting $\bb v=(0.02, 0,01)^\intercal$ and the pick matrix 
$$\bb P= \begin{pmatrix} 1 &0 &-1 &0 & ...& 0\\
0 &1 &0 &-1 & ...& 0 \end{pmatrix}. $$ 
Uncertainty is captured by assuming that $\bb v$ is subject to normally distributed perturbations $\bbb \varepsilon$. Thus   $\bb v$  is a random variable $\bb V$ such that $\bb V=\bb P\bbb \mu +\bbb \varepsilon$ with $\bbb \varepsilon \sim\tn N(\bb 0,  \bbb \Omega)$ and $\bbb \mu, \bbb \varepsilon$ independent.

\cite{CeHaMePfe2021JBF}  consider an extended BL-model denoted by {\sl EBL}. Instead of using quantitative views, the authors process qualitative views which simply express ordering relations between asset returns.
These are represented as inequalities   $\bb P\bbb \mu \geq \bb 0$, where $\bb P$  is again a  pick matrix.
For example, the qualitative view that asset 1 will outperform asset 3 and asset 2 will outperform asset 4 can be modeled by the pick matrix 
$$\bb P= \begin{pmatrix} 1 &0 &-1 &0 & ...& 0\\
0 &1 &0 &-1 & ...& 0 \end{pmatrix}. $$ 
The uncertainty is captured by  the random variable $\bb V$ fulfilling  $\bb V=\bb P\bbb \mu+\bbb \varepsilon$ and $\b V\ge 0$ with $\bbb \varepsilon \sim\tn N(\bb 0,  \bbb \Omega)$ and $\bbb \mu, \bbb \varepsilon$ independent.
The conditional expectation of $\bbb\mu$ given the event $\bb V \geq \bb 0 $ will be denoted by $\bbb\mu_{\tn{EBL}}$
and serves as the input to a mean-variance portfolio optimization problem.

A frequent choice for the uncertainty $\bbb\Omega$ is a matrix proportional to
the covariance matrix of $\bb P \bbb \mu$  
(see e.g.~\cite{walters2014black}), i.e.\
\begin{equation}\label{omegachoice}
\bbb\Omega=c\, \bb P\bbb\Sigma\bb P^\intercal,
\end{equation}
where the parameter $c$ represents an overall confidence level in the views with $c=0$ corresponding to full confidence in the views.
Under this setting \cite{CeHaMePfe2021JBF} derive the following expression for $\bbb\mu_{\tn{EBL}}$:
\begin{equation}\label{muEBL}
\bbb\mu_{\tn{EBL}} =  \bbb\pi + \frac{\tau}{\tau + c}\bbb\Sigma\bb P^\intercal \left(\bb P\bbb\Sigma \bb P^\intercal  \right)^{-1} (\tn E(\bb V \mid \bb V \geq \bb 0 ) - \bb P\bbb\pi)
\end{equation}  
For the numerical computation of  $\bbb\mu_{\tn{EBL}}$ 
according to (\ref{muEBL}) the authors apply an an {\em Importance Sampling} approach.

\medskip
\cite{CeHaMePfe2021JBF} provide us with an algorithmic framework which outputs a vector $\bbb\mu_{\tn{EBL}}$
of expected return estimators combining information from historical data with a view on the future development represented by order relations on the expected future asset returns.
The outcome $\bbb\mu_{\tn{EBL}}$ can finally be used as an input for the following classical mean-variance portfolio optimization model (\cite{Markowitz_1952})
with a risk-aversion parameter $\delta > 0$: 
\begin{align}
\label{MVO3}
\max_{\bb w}\ \ &\bb w^\intercal\bbb \mu - \frac{\delta}{2 } \bb w^\intercal  \bbb \Sigma \bb w \\
\tn{s.t.}\ \ & \sum_{i=1}^n w_i = 1\ \tn{and}\ w_i \geq 0\ \ (i=1,\ldots,n) \notag
\end{align}

While the approach of \cite{CeHaMePfe2021JBF} allows to account for qualitative information in portfolio optimization, the restriction of views to one single total order is a major loss of potentially relevant information.
Rather, it is reasonable to assume that investors consider multiple views (e.g., from different analysts) when constructing  their optimal portfolios. 
Ín the following, we consider \textit{multiple} views represented  by $K$ different total orders of expected asset returns. 
Formally we use $K$ permutations $\sigma_k$, $k=1,\ldots,K$,
each of them describing a total order of expected asset returns.

We compare two different approaches to incorporate these views into portfolio optimization. In Section~\ref{sec:robust} we first apply the extended BL approach  described above to obtain  a return estimator $\bbb\mu_k$ for each permutation $\sigma_k$. Thus, $\bbb\mu_k$ is obtained by applying (\ref{muEBL}), where $\bb P$ is replaced by the pick matrix $\bb P_k$  corresponding to permutation $\sigma_k$.
Then, we take the vectors $\bbb\mu_k$, $1\le k\le K$ as possible scenarios for $\bbb\mu$ in a robust version of  (\ref{MVO3}) with the discrete uncertainty set $\{\bbb\mu_k\colon 1\le k\le K\}$.
A completely different approach is pursued in Section~\ref{sec:social}, where we apply aggregation rules from social choice theory to reach a single total order  representing a consensus over the  given permutations $\sigma_k$, $1\le k\le K$.
Then, according to the extended BL approach,  we use (\ref{muEBL}) to compute a return estimator $\bar{\bbb\mu}$ for the consensus total order. Finally, the MVO  (\ref{MVO3}) is solved with $\bar{\bbb\mu}$  as an input.

\section{Robust optimization with discrete scenarios}
\label{sec:robust}

Robust optimization (RO), introduced by Ben-Tal and Nemirovski in the late 1990s in the context of linear and convex optimization problems, is a methodology for dealing with optimization problems whose input  has some uncertainty in the form of  deterministic variability on the data (for a comprehensive survey, we refer to \cite{BenTalNemi2002}).  
The possible values of a parameter or a set of  parameters are called scenarios. 
The scenarios represent possible realizations of data and take their values from a discrete uncertainty set or an uncertainty continuum.  
The goal is to find  solutions that are of  high  quality for each scenario or realization of the data. 
There are different ways to define the quality of a solution with respect to all scenarios, and different definitions of solution quality lead to different robust optimization models.

In this paper we consider the classical mean-variance portfolio optimization model (\ref{MVO3}), where   the input parameter $\bbb\mu$ is uncertain and its realizations build a discrete uncertainty set ${\cal U}:=\left \{\bbb\mu_k\colon k\in \{1,2,\ldots,K\} \right \}$.  
We assume that there are $K$ qualitative views, each of which is specified as a total order $\sigma_k$  of the assets in the  portfolio  according to the magnitude of their expected returns, $k\in \{1,2,\ldots,K\}$.  
For any $k$ we  apply the algorithmic framework of  \cite{CeHaMePfe2021JBF}  to  incorporate the $k$-th total order $\sigma_k$ in the   vector  $\bbb\mu_k$ of  expected asset returns. 
In particular $\bbb\mu_k$ is obtained by applying \eqref{muEBL} with a pick matrix $\bb P_k$  which represents the 
total order of expected asset returns specified by the permutation $\sigma_k$, for all $k\in \{1,2,\ldots,K\}$:
\begin{equation}
\label{mu_k}
\bbb\mu_k:=  \bbb\pi + \frac{\tau}{\tau + c}\bbb\Sigma\bb P_k^\intercal \left(\bb P_k\bbb\Sigma \bb P_k^\intercal  \right)^{-1} (\tn E(\bb V \mid \bb V\geq \bb 0 ) - \bb P_k\bbb\pi).
\end{equation} 
In the following we briefly discuss the three  robust optimization variants  of  (\ref{MVO3}) addressed in this paper:
max-min robustness, min-max regret robustness and soft robustness. 

\subsection{Max-min robustness}
\label{sec:maxmin}

In max-min robustness, also called the strictly robust optimization approach, see \cite{BenTalNemi2002}, the goal is to find  a solution which maximizes the worst value of the objective function over all scenarios in ${\cal U}$. 
 \begin{align}
\label{robopt1}
\max_{\bb w}\  \min_{\bbb\mu}\ \ &\bb w^\intercal\bbb \mu - \frac{\delta}{2 } \bb w^\intercal  \bbb \Sigma \bb w \\ 
\tn{s.t.}\  &\bbb\mu\in \{ \bbb\mu_1,\ldots, \bbb\mu_K\}\notag\\
&\sum_{i=1}^n w_i = 1\ \tn{and}\ w_i \geq 0\ \ (i=1,\ldots,n). \notag
\end{align}
It is obvious that this robust version of the mean-variance formulation is very conservative; a candidate portfolio $\bb w$ is considered to be optimal if it maximizes the worst case, i.e.\  the  smallest value of the risk adjusted expected return that this portfolio achieves over all  possible realizations of expected asset returns in ${\cal U}$.  
The above problem can be reformulated equivalently as a concave quadratic maximization problem. 
\begin{align}
\label{robopt2}
\max_{\bb w, y}\ \  & y - \frac{\delta}{2 } \bb w^\intercal  \bbb \Sigma \bb w \\ 
\tn{s.t.}\  & y \leq \bbb \mu_k^\intercal\bb w \ \ (k=1,\ldots,K)   \notag \\
&\sum_{i=1}^n w_i = 1\ \tn{and}\ w_i \geq 0\ \ (i=1,\ldots,n). \notag
\end{align}

It is well known that concave quadratic programming, i.e.\ maximizing a concave quadratic objective function subject to linear and affine constraints is (weakly) polynomially solvable (see e.g.~\cite{Vava2001}) and efficient optimization algorithms for solving this type of problems are available in every standard optimization software. 
Thus, in this case the max-min robust problem  \eqref{robopt2} (or equivalently  \eqref{robopt1}) remains polynomially solvable and is computationally not harder than the non-robust counterpart (\ref{MVO3}). 
Note, however, that the time complexity of the algorithms for solving  \eqref{robopt2}  depends on the number $K$ of scenarios (or views).  

\subsection{Min-max regret robustness}\label{sec:regret}

The min-max  regret robustness model is less conservative than the max-min robustness model.
In this model the quality of a solution is measured in terms of the {\sl maximum regret} defined as follows.
\begin{definition}
The maximum regret $MaxReg(\bb w)$ of a portfolio $\bb w$ with respect to the mean-variance portfolio optimization model (\ref{MVO3}) and the  discrete uncertainty set 
${\cal U}:=\big \{\bbb\mu_k\colon k\in \{1,2,\ldots, K\}  \big \}$ is defined as 
\begin{equation}\label{maxregret}
MaxReg(\bb w):=\max_{\bbb \mu  \in \{\bbb \mu_1,\bbb \mu_2,\ldots,\bbb \mu_K\}} [f(\bb w_{\bbb \mu},\bbb \mu)-f(\bb w,\bbb \mu)]\, ,
\end{equation}
where $f(\bb w,\bbb \mu)$ is the objective function of (\ref{MVO3}), 
i.e.\ $f({\bb w},{\bbb \mu}):={\bb w}^\intercal {\bbb \mu}-\frac{\delta}{2 } \bb w^\intercal  \bbb \Sigma \bb w$,
and $\bb w_{\bbb\mu}$ maximizes $f(\bb w,\bbb\mu)$ for fixed $\bbb \mu$, i.e.\
\[ f(\bb w_{\bbb \mu},\bbb \mu) =\max\left \{ f(\bb w,\bbb \mu)\colon \bb w\in \rz^n_+, \sum_{i=1}^n w_i=1\right \} \, .\]
\end{definition}

\noindent The expression in the square parentheses in (\ref{maxregret}) is called  the {\sl regret} of the portfolio $\bb w$ for a given $\bbb \mu$ and is  equal to the portion of the objective function value lost when the optimal portfolio $\bb w_{\bbb \mu}$ with respect to $\bbb \mu$ is replaced by the portfolio $\bb w$.
\smallskip

The min-max  regret robustness problem is then given as 
\begin{equation} \label{MinRegP}
\min  \{ MaxReg(\bb w) \colon \bb w\in \rz^n_+, \sum_{i=1}^n w_i=1 \} \, .
\end{equation}

For $k  \in \{1,2,\ldots,K\}$ the portfolios $\bb w_{\bbb \mu_k}$ and the values $f_k:=f(\bb w_{\bbb \mu_k},\bbb \mu_k)$ of the risk-adjusted return can be computed by  setting $\bbb \mu:=\bbb \mu_k$ for $k\in \{1,2,\ldots,K\}$ and solving $K$ mean-variance portfolio optimization problems as in (\ref{MVO3}).  Observe that 
$$MaxReg(\bb w):=\max_{k \in \{1,2,\ldots,K\}}\big [f_k-\bb w^\intercal \bbb \mu_k+\frac{\delta}{2}\bb w^\intercal \bbb \Sigma \bb w\big ]=\max_{k \in \{1,2,\ldots,K\}}\big [f_k-\bb w^\intercal\bbb \mu_k\big ]+\frac{\delta}{2}\bb w^\intercal \bbb \Sigma \bb w \, .$$ 
Thus, (\ref{MinRegP}) can be  equivalently written as   
\[ -\max \left \{ - \max _{k \in \{1,2,\ldots,K\}}[f_k-\bb w^\intercal\bbb \mu_k] - \frac{\delta}{2}\bb w^\intercal \bbb \Sigma \bb w \colon \bb w\in \rz^n_+, \sum_{i=1}^n w_i=1 \right \} \, .\]
By setting $-y_{\bb w}:=\max_{k\in \{1,2,\ldots,K\}} \big [ f_k -\bb w^\intercal \bbb \mu_k\big ]$ we get the next  equivalent formulation of (\ref{MinRegP}) as 
\[ -\max \left \{ y_{\bb w}-\frac{\delta}{2} \bb w^\intercal\bbb \Sigma \bb w \,\colon -y_{\bb w}\ge f_k -\bb w^\intercal\bbb \mu_k, k\in \{1,2,\ldots,K\}, \bb w\in \rz^n_+, \sum_{i=1}^n w_i=1 \right \} \, .\] 
Finally, by observing that  in the last formulation the variable $y_{\bb w}$ can be replaced by a single variable $y$  (and by neglecting the  minus in front of the problem) we get the following concave quadratic maximization formulation of  (\ref{MinRegP})  
\begin{align}
\label{robopt3}
\max_{\bb w, y}\ \  & y - \frac{\delta}{2 } \bb w^\intercal  \bbb \Sigma \bb w \\ 
\tn{s.t.}\  & y \leq \bbb \mu_k^\intercal\bb w -f_k\, ,  \, k=1,\ldots,K   \notag \\
&\sum_{i=1}^n w_i = 1\ \tn{and}\ w_i \geq 0\, , \, i=1,\ldots,n. \notag
\end{align}

Thus, similarly to the case of the max-min robustness approach, also  the min-max  regret robustness problem  \eqref{MinRegP} (or equivalently  \eqref{robopt3}) remains polynomially solvable and is not harder than its non-robust counterpart (\ref{MVO3}). 
Note, however, that the time complexity of algorithms for solving  \eqref{robopt3}  heavily depends on the number $K$ of scenarios (or views). 
Other than in the case of the max-min robust version of the mean-variance formulation, here we also need to solve $K$ concave quadratic minimization problems to determine $f_k$, $k\in \{1,2,\ldots,K\}$, prior to solving the concave quadratic maximization problem   \eqref{robopt3} which analogously to \eqref{robopt2} involves $K+1$ linear  constraints  besides the nonnegativity constraints. 
\subsection{Soft robustness}
\label{sec:soft}

The soft robustness model allows to  control the degree of conservativeness of
the robust  solution in the vein of the so called $\Gamma$-robustness (see
\cite{BerSim2003}).
This model includes a parameter $\Gamma\in (0,1]$ which allows to  adjust the extent at which the possible estimators of the expected asset returns $\bbb \mu_k$, $k\in \{1,2,\ldots, K\}$, are taken into account. More precisely, we want to choose a portfolio $\bb w^{\ast}$ which maximizes a threshold of risk adjusted return which would be exceeded by at least $\Gamma K$ estimators  $\bbb \mu_k$, $k\in \{1,2,\ldots, K\}$. In other words $\bb w^{\ast}$ maximizes the empirical $(1-\Gamma)$-quantile $y_{\Gamma}(\bb w)$ of the risk-adjusted expected returns over all feasible portfolios $\bb w$.
\begin{definition}
The empirical $(1-\Gamma)$-quantile $y_{\Gamma}(\bb w)$  of the set \\ $\{f(\bb w, \bbb \mu_1), f(\bb w, \bbb \mu_2),\ldots, f(\bb w, \bbb \mu_K)\}$ is defined as follows
\[ y_{\Gamma}(\bb w):= \sup \Big \{z\in \rz \colon \big\vert \big \{k: k\in \{1,2,\ldots,K\}, f(\bb w,\bbb \mu _k)\ge z\big \}\big\vert \ge \Gamma K \Big  \}\, .\] 
\end{definition}

The corresponding optimization problem is then given as 
\begin{equation} \label{robopt4}
\max \left \{ y_{\Gamma}(\bb w)\colon \bb w\in \rz^n_+, \sum_{i=1}^n w_i=1\right \}\, .
\end{equation}

By introducing a new variable $y$ to represent the maximum of $y_{\Gamma}(\bb w)$  we obtain the following equivalent formulation  of (\ref{robopt4}).

\begin{align}
\label{robopt5}
\max_{\bb w, y}\ \  & y \\ 
\tn{s.t.}\  &  \big\vert \{k: k\in \{1,2,\ldots,K\}, f(\bb w,\bbb \mu _k)\ge y\}\big\vert \ge \Gamma K \notag \\
&\sum_{i=1}^n w_i = 1\ \tn{and}\ w_i \geq 0\, , \,  i=1,\ldots,n. \notag
\end{align}
Obviously, the larger   the value of $\Gamma$,  the more conservative is the model. In the limit case $\Gamma=1$ we get  $y_{\Gamma}(\bb w)= \min \{f(\bb w,\bbb \mu_1), f(\bb w,\bbb \mu_2),\ldots, f(\bb w,\bbb \mu_K)\}$  and the soft robustness model coincides with max-min robustness model. On the other hand,  if  $\Gamma$ becomes very small, e.g.\ $\Gamma=\frac{1}{K}$,  we get   $y_{\Gamma}(\bb w)= \max \{f(\bb w,\bbb \mu_1), f(\bb w,\bbb \mu_2),\ldots, f(\bb w,\bbb \mu_K)\}$  and the soft robustness model becomes 
\begin{align}
\notag
\max_{\bb w} \max_{\bbb \mu}\ \ & \bb w^\intercal \bbb \mu - \frac{\delta}{2} \bb w^\intercal \bbb \Sigma\bb w \\ 
\tn{s.t.}\  &  \bbb \mu \in \{\bbb \mu_1,\bbb \mu_2,\ldots,\bbb \mu_K\}\notag \\
&\sum_{i=1}^n w_i = 1\ \tn{and}\ w_i \geq 0\, , \,  i=1,\ldots,n. \notag
\end{align}
\smallskip

Observe now that (\ref{robopt5}) can be formulated as a convex  mixed integer nonlinear problem  (MINLP) whith a linear objective function involving just continuous variables as well as linear and convex quadratic constraints. Indeed, we introduce the binary variables $v_k\in \{0,1\}$, $k \in \{1,2,\ldots,K\}$,   such that $v_k=0$ iff $y \le f(\bb w,\bbb \mu_k)$ and $v_k=1$, otherwise. In the latter case $y\le f(\bb w, \bbb \mu_k)+v_kM$ is  fulfilled  for a large enough  constant $M$. The constraint $\sum_{k=1}^K v_k\le (1-\Gamma)K$ guarantees that at least $\Gamma K$ of the variables $v_k$ equal zero, i.e.\ that $f(\bb w,\bbb \mu_k)\ge y$ holds for at least $\Gamma K$ estimators $\bbb \mu_k$. Finally, by introducing an additional auxiliary variable $z$ as an upper bound on $\bb w^\intercal \Sigma \bb w$ the inequality $y\le f(\bb w, \bbb \mu_k)+v_kM$ can be formulated as a linear constraint $y\le \bb w^\intercal \bbb \mu_k-\frac{\delta}{2}z+v_k M$. Summarizing we obtain the following equivalent formulation of (\ref{robopt5}):
\begin{align}
\label{robopt6}
\max_{\bb w} \ \ & y \\ 
\tn{s.t.}\  &  y\le \bb w^\intercal \bbb \mu_k -\frac{\delta}{2} z +v_k M \, , \, k =1,2,\ldots,K \notag \\
& \bb w^\intercal \bbb \Sigma \bb w  \le z\notag \\
& \sum_{k=1}^n v_k \le (1-\Gamma)K\notag \\
&\sum_{i=1}^n w_i = 1\ \tn{and}\ w_i \geq 0\, ,  \, i=1,\ldots,n\notag\\
& v_k\in \{0,1\}\, , \,  k=1,\ldots,K. \notag 
\end{align}

In general the convex  MINLP is a NP-hard problem because it includes the  mixed integer linear programming (MILP), see e.g.\ \cite{Belottietal2013}. There is a wealth of literature on exact and heuristic solution approaches for convex MINLP, see e.g.\ the recent  review by \cite{Kronqvistetal2019}. 
Note, however, that (\ref{robopt6}) has a particular structure: the constraints involving the binary variables are linear and the objective function  just involves continuous variables.
The outer approximation method of \cite{DuranGrossmann86} is an iterative approach  which exploits  the particular structure mentioned above and consists in solving an  alternate  finite sequence of convex quadratic subproblems and relaxations of a MILP as a master problem.
A rigorous analysis of the convergence properties of the algorithm can also be
found in \cite{Belottietal2013}.


\section{Aggregation by social choice methods}
\label{sec:social}

In this section we pursue the idea of computing a single consensus total order of expected asset returns which is a sort of ``center'' of all $K$ given total orders.
Formally, we are looking for a total order $\sigma^*$ which minimizes the sum of distances to the total orders $\sigma_1, \ldots, \sigma_K$
(see \cite{marden96} for an overview).
The distance between two total orders $\sigma'$ and $\sigma''$ will be measured by the classical Kendall-Tau distance which counts the number of pairwise disagreements between two total orders,
i.e.\ the number of pairs $(i,j)$ with $\sigma'(i) < \sigma'(j)$ and
$\sigma''(i) > \sigma''(j)$. For our purposes we use the normalized version of the Kendall-Tau distance  $KT(\sigma',\sigma'')$ given as the number of pairwise disagreements between $\sigma'$ and $\sigma''$  divided by the overall  number of pairs $(i,j)$ with $i\neq j$, see e.g.~\cite{dwork2001rank}.
Then we are looking for an optimal total order $\sigma^*$ such that
\begin{equation}\label{eq:pistar}
    \sigma^* = \arg \min_{\sigma\in S_n} \sum_{k=1}^K KT(\sigma, \sigma_k),
\end{equation}
where $S_n$ is the set of all permutations of numbers $1,\ldots,n$.
This optimization problem, which is called the Kemeny-Young
problem,
is the basis for the well-known Kemeny-Young method for group decisions.
However, the problem is NP-hard  for $K\geq 4$ (\cite{dwork2001rank}) and becomes practically intractable already for a moderate number of assets.
Thus, we will focus on approximate solutions to the Kemeny-Young problem, in particular on methods developed in the area of social choice.

\medskip
Social choice theory (see e.g.\ \cite{socialhand02}) studies the aggregation of individual preferences or opinions into a group decision considering issues such as fairness, social welfare or acceptability.
Typical areas of application are voting, selection of committees and allocation of resources. 
A central concept in this field is a {\em social welfare function} (cf.\ the seminal work of Nobel price winner \cite{collective70}),
which receives as input $K$ total orders of the objects $1,\ldots,n$ representing the preferences of $K$ individuals and determines as output a single total order.
This is exactly the task we are facing in this section.
The mapping of the total orders $\sigma_k$, $1\le k\le K$, to $\sigma^*$  in (\ref{eq:pistar}) can be seen as one particular social welfare function. For a total order $\sigma$ we  call the sum in the righthand side of (\ref{eq:pistar}) the Kendall-Tau score of $\sigma$ (with respect to the input total orders $\sigma_k$, $1\le k\le K$).

In social choice theory a large number of social welfare functions were introduced and their theoretical properties investigated.
A good overview can be found in \cite{brfi02}.
In the following we will describe some of the more prominent examples which will then be applied to compute a total order of expected asset returns.
For a broader discussion and computational aspects we refer the reader to \cite{Brandt16}.

\subsection{Borda rule}

The Borda rule was formulated by Jean-Charles Chevalier de Borda in 1770 and is one of the most prominent and natural aggregation rules.
It belongs to the family of scoring rules, where an element at a certain position of a total order is assigned a certain number of points.
In case of the Borda rule, the asset $i$ at position $\sigma(i)$ in a total order gets a Borda score $Bor_\sigma(i):=n-\sigma(i)$.
The aggregation works by simply summing up for every asset $i$ the scores over all $K$ total orders, i.e.\ $\sum_{k=1}^K Bor_{\sigma_k}(i)$, and sorting the assets in decreasing order of these total scores.
\subsection{Footrule Aggregation}

This aggregation is based on an optimization problem similar to (\ref{eq:pistar}),
but it applies the \emph{Spearman footrule} $SF(\sigma', \sigma'')$ for measuring the distance between two total orders $\sigma'$ and $\sigma''$ instead of the Kendall-Tau distance.
It is based on measuring for every $i$ the absolute difference between the positions $\sigma'(i)$ and $\sigma''(i)$  and summing up these values. 
Formally, we are looking for a total  order $\sigma^*$ such that
\begin{equation}\label{eq:footrule}
    \sigma^* = \arg \min_{\sigma\in S_n} \sum_{k=1}^K SF(\sigma, \sigma_k)\,, \quad \mbox{ with }
    SF(\sigma, \sigma_k) = \sum_{i=1}^n \vert\sigma(i) - \sigma_k(i)\vert.
\end{equation}
It was shown by \cite{diaconis1977spearman} that the difference between the Spearman footrule distance and the  Kendall-Tau distance can be bounded by
\begin{equation}\label{eq:footrulecomp}
KT(\sigma', \sigma'') \leq SF(\sigma', \sigma'') \leq 2 \,KT(\sigma', \sigma'').
\end{equation}
Moreover, $SF(\sigma', \sigma'')$ can be computed efficiently in polynomial time by solving a linear assignment problem as shown e.g.\ in \cite{dwork2001rank}.
\subsection{Copeland Method}

For a given set of $K$ total orders, we introduce a relation $\succsim$ (majority dominance) on the set of assets such that $i \succsim j$ if and only if $i$ is ranked higher than $j$ by a majority of the total orders, i.e.\
$\vert\{k\in\{1,\ldots, K\} : \sigma_k(i) < \sigma_k(j)\}\vert \geq K/2$,
see e.g.~\cite{fis77}.
The Copeland value $Cop(i)$ of an asset $i$ is defined as the number of assets majority dominated by $i$ minus the number of assets majority dominating $i$, i.e.\
$$Cop(i):= \vert\{j\in \{1,\ldots,n\} : i\succsim j\}\vert - 
 \vert\{j\in \{1,\ldots,n\} : j\succsim i\}\vert$$
For use as an aggregation rule, assets are ordered in decreasing order of their Copeland values.
\subsection{Best-of-$k$-Algorithm}

This simple approach solves the Kemeny--Young problem under the strong restriction that $\sigma^*$ is chosen only among the given total orders, 
i.e.\ in (\ref{eq:pistar}) the domain of the minimization $\sigma\in S_n$
is replaced by $\sigma\in\{\sigma_1, \ldots, \sigma_K\}$.
Thus, it suffices to calculate the Kendall-Tau distance between all pairs of total orders and then compare the appropriate sums.
This straightforward approach yields a $2$-approximate solution to (\ref{eq:pistar}) with a Kendall-Tau distance at most twice as large as given by $\sigma^*$ (see e.g.~\cite{ailon2008aggregating}).

\subsection{MC4-Algorithm}

In their widely cited paper \cite{dwork2001rank} present four aggregation algorithms based on Markov chains.
Among these, we will consider the so-called MC4-Algorithm, which performed best in the empirical evaluation of~\cite{dwork2001rank} and~\cite{schalekamp2009rank}.
In this approach, the $n$ assets correspond to the $n$ states of a Markov chain.
Starting from a randomly chosen state, the Markov process moves in each iteration from the current state to a new state of the system. 
From a current state $i$, the new state $j$ is chosen with a certain transition probability $p_{ij}$.
If the current situation of the system is represented by a probability distribution $\bbb x$, where $x_i$ gives the probability that the process is currently in state $i$,
then the probability distribution of the next state is given by $\bbb x^\intercal \bbb P$, where the probability matrix $\bbb P =(p_{ij})$ consists of the transition probabilities.
It is known that under certain conditions, this process reaches a stationary distribution $\bbb y$ with $\bbb y = \bbb y^\intercal \bbb P $, and $\bbb y$ can be computed e.g.\  by a simple power-iteration algorithm.

When applying MC4  in the aggregation of total orders, the probabilities $p_{ij}$ are defined as follows:
if $i$ is the current state, a potential new state $j$ is chosen from $\{1,\ldots, n\}$ uniformly at random. 
If a majority of the total orders ranks $j$ before $i$,
i.e.\  $j \succsim i$ according to the majority dominance of the Copeland Method\footnote{Ties are broken lexicographically, i.e.\ by the index number of the assets.}, then the Markov process moves to state~$j$, otherwise it remains in state~$i$.
To ensure that the process has a unique stationary distribution \cite{schalekamp2009rank} add a random ``jump'': in every step the process moves from $i$ to the randomly chosen $j$ with a tiny probability $\alpha$, independently from the dominance relation. 
With probability $1-\alpha$, the move to $j$ is only chosen if $j \succsim i$.
Hence, there is a transition probability $p_{ij}=1/n + \alpha/n$, if $j \succsim i$, and $p_{ij}=\alpha/n$, otherwise.
The non-transition probabilities are then set  as $p_{ii}=1- \sum_{j\neq i} p_{ij}$, for all $i$.
We use $\alpha=1/100$ to ensure that for up to $n=100$ assets the sum of the transition probabilities does not exceed $1$.

After reaching the unique stationary distribution $\bbb y$,
the aggregation of total orders follows by sorting the assets in decreasing order of $y_i$.

\subsection{Local Improvement}

Any total  order $\sigma$ produced by one of the above aggregation methods can be further improved 
by reducing the corresponding value of the objective function in  (\ref{eq:pistar}), i.e.\ the Kendall-Tau score of $\sigma$.
In our approach this is done by an iterative pairwise exchange operation.
We consider the assets in increasing order, i.e.\ we go through all positions $\ell=2,3, \ldots, n$. In the  iteration for asset $i$ in position $\ell$,  $\sigma(i)=\ell$,
we  compare $i$ with the asset $j$ at the preceding position $\sigma(j)=\ell-1$. 
If $j\succsim i$ (according to the majority dominance), then and we proceed to the next iteration for the asset in  position $\ell+1$.
Otherwise, if $i\succsim j$, we exchange $i$ and $j$ in $\sigma$ by setting $\sigma(i)=\ell-1$ and $\sigma(j)=\ell$, thereby improving the Kendall-Tau score of $\sigma$.
We continue  by  comparing  $i$ with the asset at position $\ell-2$ and so on, until either  a comparison without improvement is found  or $i$ reaches the first position. 
At this point the iteration for asset $i$ terminates and we move to the next iteration for the immediate successor of asset $i$ at position $\ell+1$.
 This procedure  is  called ``Local Kemenization'' in \cite{dwork2001rank} and ``InsertionSort'' in \cite{schalekamp2009rank}. 

In our experiments we apply this process of \emph{Local Improvement} to all total orders produced by the above aggregation procedures.
\section{Computational experiments}
\label{sec:comp}
In this section, we evaluate the performance of the different approaches presented in Sections~\ref{sec:robust} and \ref{sec:social} for aggregating multiple qualitative views, expressed as total orders of expected returns of assets in the portfolio, in the context of classical MVO using real stock market data.
 We apply the extended BL-model suggested in   \cite{CeHaMePfe2021JBF}, more precisely the equality (\ref{muEBL}), to compute an estimator of expected asset returns  for any given total order of the latter (see Section~\ref{sec:formal}).

We use historical data from two stock indices, one in Europe and one in the U.S. (see Section~\ref{sec:data}).
For both samples, we compare the performance of  monthly rebalanced MVO portfolios based on (I) the several robust optimization approaches and (II) the various ordering aggregation methods based on  social choice theory. 

In a first step, we compare the performance of the different approaches within each group separately. 
To mitigate the effect of estimation errors   the analysis is based on  in-sample tests. The tests are in-sample in  that (a)  the complete time series of realized asset returns are used to   estimate the corresponding covariance matrix (which then remains constant over the entire time horizon) and (b) the synthetic views (i.e.\ total orders) on  the  expected asset returns in the time interval $(t,t+1)$
are randomly generated (see below) utilizing information of the actual  total order of asset returns realized in that time interval. 
{We are aware that our procedure for generating ordering relations relies on a theoretical setting. One can imagine that in practice these relations could be derived, for example, from analysts' earnings estimates. Most likely, however, these are not total orders since financial analysts can at best make such forecasts for a limited number of individual stocks in a given industry or for a few different asset classes. Alternatively, rankings could be created by using some technical indicators (such as the Moving Average Convergence Divergence, the Bollinger Bands or any Fibonacci retracement levels; see for example \cite{tech_anal}), which could be  converted into total orders.  
Similarly, one can determine the intrinsic value of a stock (e.g., by combining financial statements, external influences, or industry trends) and rank stocks according to the difference between their current market prices and their intrinsic values. 
However, the aim of our analysis is not to present ways to generate ordering relations between asset returns, but rather to present different methods for aggregating multiple views and to compare the performance of these approaches given different ordering relations in a MV analysis. Therefore, we separate the process of generating rankings from the process of processing these rankings and focus on the latter.}
 
In a second step and after identifying the two best performing approaches  from each group  (I) and (II), we  compare the performance of the two winners of each group to identify the overall ``champion''.

 All experiments were implemented and run in Matlab R2021b on a standard PC.\footnote{Since the computations required negligible running times and the computational performance was not the focus of this study we refrain from describing further technical details.}

\subsection{Data description}
\label{sec:data}
We create two samples with single stock data from the EURO STOXX 50 and the S\&P 100. 
The EURO STOXX 50 is a blue-chip stock index that comprises fifty of the largest and most liquid stocks in the Eurozone. 
The S\&P 100, a subset of the S\&P 500, includes the largest U.S.\ companies across a variety of industries. 
We collect monthly total return index data from Refinitiv Datastream for the period December 1998 to December 2021 for all stocks included in either index at the end of 2021. 
We eliminate stocks for which we do not have a continuous price series over the entire time period. 
In this way we identify 38 (76) stocks from EURO STOXX 50 (S\&P 100) that are eligible for our investigation.
We calculate monthly returns from the total return index series, with January 1999 being the first observation, giving a total of 276 monthly returns for each stock.  
Table \ref{tab:descript} reports descriptive statistics
for cross-sectional averages for our sample of constituents of the EURO STOXX 50 (Panel A) and the S\&P 100 (Panel B). For the former, the average monthly holding period return (in EUR) is 1.05\% with an average standard deviation of 8.33\%. 
The corresponding values for our S\&P 100 sample are 1.14\% for the average monthly return (in USD) and 8.21\% for the standard deviation. Concerning skewness, we find evidence of slightly positive values, while kurtosis shows the typical pattern of fat tails in both samples.

\begin{table}[h]
	\footnotesize 
	\centering

\begin{tabular}{lrrrrrr}
		

		\multicolumn{7}{l}{Panel A: Selected constituents of EURO STOXX 50}       \\ \hline
		
		monthly return (\%) & \multicolumn{1}{c}{mean} & \multicolumn{1}{c}{q1} & \multicolumn{1}{c}{q2} & \multicolumn{1}{c}{q3}  & \multicolumn{1}{c}{min}  & \multicolumn{1}{c}{max}  \\ \hline
		mean        & 1.05    & 0.76 & 0.96 & 1.22 & 0.42 & 2.32  \\
		stdev       & 8.33    & 6.62 & 8.57 & 9.52 & 4.79 & 11.79   \\
		skew        & 0.15   & -0.13 & 0.13 & 0.32 & -1.02 & 1.22  \\
		excess kurtosis	& 3.74    & 1.10 & 1.90 & 4.74 & 0.34 & 20.30	\\\hline
		&         &               &        &        &        &        \\   
		\multicolumn{7}{l}{Panel B: Selected constituents of S\&P 100}            \\ \hline
		
		monthly return (\%) & \multicolumn{1}{c}{mean} & \multicolumn{1}{c}{q1} & \multicolumn{1}{c}{q2} & \multicolumn{1}{c}{q3}  & \multicolumn{1}{c}{min}  & \multicolumn{1}{c}{max}  \\ \hline
		mean        & 1.14    & 0.81 & 1.02 & 1.37 & 0.22 & 3.03  \\
		stdev       & 8.21    & 6.51 & 7.46 & 9.12 & 4.73 & 20.81   \\
		skew        & 0.31   & -0.23 & 0.08 & 0.38 & -1.38 & 6.51  \\
		excess kurtosis	& 5.82    & 1.53 & 2.64 & 4.84 & 0.21 & 76.35	\\\hline
	\end{tabular}

\caption{Descriptive statistics for monthly returns (in \%) of 38 stocks in the EURO STOXX 50 (Panel A) and 76 stocks in the S\&P 100 (Panel B).
All stocks are included a) that were constituents of one of the corresponding stock indices on December 31, 2021, and b) for which a
complete return time series is available for the period from January 1999 to December 2021. The values for mean and stdev are given in \%. q1 (q2) [q3] 
represents the first (second) [third] quartile. All measures are expressed as
equally weighted cross sectional averages. Monthly returns are 
calculated as holding period returns from the last day of one month to the last
day of the following month.  Sample period: January 
1999 to December 2021. Source: Refinitiv Datastream.}
\label{tab:descript}
\end{table}

\subsection{Setup of the experiments}
\label{sec:compsetup}

We choose a large number of different parameter settings for our experiments. 
For each of the two samples we generate $K$ total orders of all assets 
for  $K\in \{5,10, 20\}$.  The  total orders represent  different synthetic  qualitative views, as they might be expressed by  different experts (e.g.\ financial analysts)  under different scenarios. We consider our choice of values for $K$ to be reasonable, but emphasize that our methods can in principle be applied to any number of views. 
We further differentiate the views according to the validity of the information
they contain. We quantify the degree of validity of information in a total
order ${\cal O}$ in terms of the Kendall-Tau 
distance $d$ of ${\cal O}$ from  the  total order of the   assets returns realized in the time interval $(t,t+1)$ which  represents the correct ordering.  Note that if the  Kendall-Tau distance between  ${\cal O}$  and the correct ordering is about $0.5$, then  ${\cal O}$ coincides with the correct ordering for about half of the pairs of assets. 
Such a total order  ${\cal O}$ would therefore represent a complete lack of knowledge about the correct ordering.  
On the other hand, a  value of $d$ around $0$ (and symmetrically a value of $d$ around $1$) corresponds to a high degree of useful information, because in this case ${\cal O}$ would coincide with the correct ordering for almost all pairs (no pairs)  of  assets. In order to cover the whole range of the possible values of  $d$ we   consider $d \in \{0.2, 0.3, 0.4, 0.47\}$. 
For each value of 
$d$ we generate $K$ random total orders of assets, each of them having the same Kendall-Tau distance $d$ from the correct ordering.\footnote{Note that we could also handle individual values of $d$ for each total order to account for differences in the quality of the views.
This can be useful, for example, when several financial analysts have different track records in terms of their forecasts and stock recommendations in the past. 
In this case the view of an analyst with a high track record could be assigned a lower value of $d$, while the view of an analyst with a poor record could be given a higher $d$.}

Finally, we consider four different levels of overall confidence in the validity of the generated total orders, represented by the  parameter $c$ (see equation (\ref{omegachoice})). Note that the same confidence level applies to each of the $K$ total orders. This should reflect the fact that it is indeed difficult to a priori assess the information content of a (qualitative) view.\footnote{Similarly to  parameter $d$ our model is fully flexible to account for different confidence levels $c$ for each total order.} We span a wide range of relevant values with $c\in\{0.25, 0.5, 0.75, 0.95\}$, where $c=0.25$ ($c=0.95$) indicates a very high (low) degree of confidence. As in \cite{CeHaMePfe2021JBF}, we set $\tau=1-c$, where the parameter $\tau$   represents the confidence in the prior information.

For each value $K\in \{5,10,20\}$, we perform our experiments for all 16 value combinations of  the parameters  $d$ and $c$, giving a total number of $48$ parameter settings.
One might argue that some parameter combinations, such as large distances and high levels of confidence, are less meaningful, but we chose to retain all combinations because it is in general unpredictable where the ordering information will come from and in what context our method will be applied.

\subsection{Measuring portfolio performance}
\label{sec:perfmeas}

Our goal is to find the best out of the aforementioned approaches that allow for integrating  multiple input orderings into MVO (cf.\ (I) and (II) at the beginning of this section). To make a decision in this ''horse race''  we compare the performance of the optimal portfolios derived from each approach by means of two metrics:\footnote{For a detailed description of the two metrics we refer to \cite{Bodieetal}. They are used, among others, in studies by \cite{DeMiguel2009a} and \cite{Allaj2020}.}




\begin{itemize}
 \item{\textit{Sharpe Ratio (SR)}:
 The Sharpe Ratio is defined as the excess return of a portfolio divided by its total risk, where the latter is  measured by the standard deviation of  portfolio returns. The excess  return is calculated with respect to  the average risk-free rate. 
  Thus, if the return and the volatility of a  porfolio $P$  are given by $\mu_P$ and $\sigma_P$, respectively, then the Sharpe Ratio $SR_P$ of the portfolio $P$ is given as\footnote{Note that since we are using in-sample mean and variance estimates, we calculate the in-sample Sharpe Ratio with portfolio weights obtained with these estimates.}  $$SR_P=\frac{\mu_P-r_f}{\sigma_P}\, , $$ where $r_f$ is the risk-free rate. We use the 3m Euribor for the EURO STOXX 50 sample and the 3m USD Libor for the S\&P 100 sample. 
  The higher \textit{SR} the higher the risk-adjusted performance of a portfolio.
  } 
 \item \textit{Certainty-Equivalent return (CEQ)}:
 The Certainty-Equivalent return is the zero-risk return that an investor with risk aversion $\delta$  is willing to accept rather than pursuing a particular risky investment strategy with a higher but riskier return. 
 Formally, the Certainty-Equivalent return $CEQ_P$ of a portfolio $P$ with return $\mu_P$ and volatility $\sigma_P$ is given as $$CEQ_P=\mu_P-\frac{\delta}{2}\sigma_P^2\, . $$ 
We set $\delta$ equal to three which is a commonly used value in the literature (e.g., \cite{CeHaMePfe2021JBF}).
 As with  \textit{SR} investors prefer portfolios with higher \textit{CEQ} values to portfolios with lower values.
\end{itemize}

\subsection{Computational comparison}
\label{compcomp}
As mentioned at the beginning of this section, we first evaluate the performance of the robust optimization-based approaches (I) and the performance of the social choice-based  approaches (II) separately and try to identify the best performing methods in each of the two groups.

In (I) we  consider six robust optimization variants: Max-min robustness (Max-min), Min-max regret robustness (Min regret) and soft robustness (Soft $\Gamma$) with $\Gamma \in \{0.25, 0.50, 0.75, 1\}$ as described in Section~\ref{sec:robust}.
In (II), we consider five ordering aggregation approaches: Borda rule (Borda), Footrule aggregation (Footrule), Best-of-$k$ algorithm (Best-of-$k$), Copeland method (Copeland)  and MC4-algorithm (Markov). As described in Section~\ref{sec:compsetup}, we conduct experiments for 48 parameter settings. 

For each group of approaches (I) or (II) and for each parameter  setting, we determine  the winners in terms of  \textit{SR} and  \textit{CEQ}. 
We define as winners all methods with a  \textit{SR}, or  \textit{CEQ} value within $1\%$ of the respective maximum value.
For each method we count the number of wins out of the 48 ``races'' and report results in Table~\ref{tab:winners}.

\begin{table}[]
	\footnotesize 
	\centering
\setlength\tabcolsep{3pt}
\begin{tabular}{l|cccc|cccc}


\multicolumn{9}{l}{Panel A: Selected constituents of EURO STOXX 50}       \\ \hline
& \multicolumn{4}{c|}{\textit{SR}}&\multicolumn{4}{c}{Certainty-Equivalent return \textit{CEQ}} \\ 
Method & $K=5$ & $K=10$ & $K=20$ &  total & $K=5$ & $K=10$ & $K=20$ &  total\\ \hline
Max-min & 3 & 3 & 2 & 8 & 3 & 2 & 1 & 6 \\
Min regret & 7 & 2 & 3 & \dbox{ 12 } & 3 & 2 & 1 & 6 \\
Soft $0.25$  & 3 & 4 & 4 & 11 & 3 & 1 & 2 & 6\\
Soft $0.5$   & 4 & 6 & 6 & \fbox{ 16 } & 9 & 10 & 7 & \fbox{ 26 } \\
Soft $0.75$  & 1 & 1 & 6 & 8 & 1 & 2 & 7 & \dbox{ 10 } \\
Soft $1$ & 1 & 1 & 0 & 2 & 1 & 2 & 0 & 3 \\\hline
Borda & 14 & 10 & 14 & \fbox{ 38 } & 15 & 13 & 14 & \fbox{ 42 } \\
Footrule & 1 & 5 & 8 & 14 & 3 & 3 & 9 & 15 \\
Copeland & 5 & 6 & 14 & \dbox{ 25 } & 6 & 12 & 15 & \dbox{ 33 } \\
Best-of-k & 3 & 0 & 8 & 11 & 3 & 2 & 9 & 14 \\
Markov & 3 & 5 & 12 & 20 & 6 & 8 & 10 & 24\\ \hline

\multicolumn{9}{c}{} \\

\multicolumn{9}{l}{Panel B: Selected constituents of S\&P 100}       \\ \hline
& \multicolumn{4}{c|}{Sharpe Ratio \textit{SR}}&\multicolumn{4}{c}{Certainty-Equivalent return \textit{CEQ}} \\ 
Method & $K=5$ & $K=10$ & $K=20$ &  total & $K=5$ & $K=10$ & $K=20$ &  total\\ \hline
Max-min & 4 & 5 & 7 & \dbox{ 16 } & 1 & 2 & 1 & 4 \\
Min regret & 9 & 6 & 7 & \fbox{ 22 } & 1 & 0 & 0 & 1 \\
Soft $0.25$  & 3 & 2 & 3 & 8 & 6 & 2 & 1 & 9\\
Soft $0.5$   & 5 & 4 & 5 & 14 & 8 & 12 & 11 & \fbox{ 31 } \\
Soft $0.75$  & 0 & 3 & 2 & 5 & 0 & 0 & 5 & 5 \\
Soft $1$ & 2 & 1 & 1 & 4 & 4 & 3 & 3 & \dbox{ 10 } \\\hline
Borda & 10 & 9 & 13 & \fbox{ 32 } & 12 & 12 & 14 & \fbox{ 38 } \\
Footrule & 2 & 8 & 8 & 18 & 4 & 6 & 9 & 19 \\
Copeland & 9 & 11 & 11 & \dbox{ 31 } & 9 & 14 & 13 & \dbox{ 36 } \\
Best-of-k & 3 & 5 & 7 & 15 & 5 & 3 & 9 & 17 \\
Markov & 9 & 6 & 9 & 24 & 6 & 10 & 12 & 28
	\\\hline	

	\end{tabular}
\caption{Number of wins of different methods of aggregating qualitative views organized in groups. Methods 'Max-min' to 'Soft 1' denote robust optimization-based  approaches  (group I), 'Borda' to 'Markov' are social choice-based approaches (group II). $K\in \{5, 10, 20\}$ indicates the number of views in terms of total orders  of assets. The input total orders  have four different values of Kendall-Tau distance $d$ to the correct order of returns, $d\in \{ 0.2, 0.3, 0.4, 0.47\}$. The level of confidence in the views $c$ takes  four different values: $c\in \{0.95,0.75,0.50,0.25\}$. The maximum number of wins for any $K$ is $16$. Results are counted as wins if they deviate from the corresponding optimum within a group  (I or II) by less than 1 percent. Solid (dashed) boxes give the highest (second highest) number of total wins within robust estimation and social choice methods, respectively. Panel A (B) shows results for 38 (76)  stocks in the EURO STOXX 50 (S\&P 100). Sample period: January 1999 to December 2021. Source: Refinitiv Datastream.}
\label{tab:winners}
\end{table}

In general, we find that there is no particular method in either (I) or (II) that clearly dominates all other methods across all parameter settings. 
We therefore select the two most successful methods from each of the groups (I) and (II). 
To this end, we consider the total number of wins over all parameter settings with respect to \textit{SR} and \textit{CEQ}, respectively. 
To illuminate the best methods within each group (I) and (II) we indicate the method with the highest (second highest) number of total wins with a solid (dashed) box in Table~\ref{tab:winners}. 

For the methods based on robust optimization, the choice of Soft $0.5$ is unambiguous: this method has the largest number of wins in $3$ out of $4$ subtables. This method is also mostly top-ranked in this group regardless of the number of views $K$. 
Other Soft methods also perform well in some races, but Soft $0.5$ is clearly the best performing variant from this family. 
As the second best method within group (I), we identify Min regret with one win and one second place out of  $4$ subtables. 
Min regret is closely followed, but not dominated, by Max-min.

Among  social choice-based  methods, Borda can be seen as the overall winner with $4$ wins and Copeland is the runner-up in all 4 subtables (even with some first places in individual settings), although the competitors are quite close. 
Therefore, and to identify the best method that allows for integrating multiple input orderings into MVO, we focus our  further  investigations on the four methods Min regret, Soft $0.5$, Borda and Copeland.

Table~\ref{tab:summary} provides more detailed results for the comparison of  these "first step winners". For each combination of the Kendall-Tau distance $d$ and the confidence level $c$, we consider the number of wins across the three portfolios obtained for the three values of $K$. These numbers are  counted separately for \textit{SR} and \textit{CEQ}.
The full table, while somewhat overwhelming at first glance, allows for relatively clear conclusions.

\begin{table}[]
\scriptsize
	\centering

\begin{tabular}{l|c|cc|cc|cc|cc}


\multicolumn{10}{l}{Panel A: Selected constituents of EURO STOXX 50}       \\ \hline
\multicolumn{2}{c|}{} &
   \multicolumn{8}{c}{Kendall tau distance} \\

 \multicolumn{2}{c|}{} &
   \multicolumn{2}{c|}{$d=0.2$}&\multicolumn{2}{c|}{$d=0.3$}&\multicolumn{2}{c|}{$d=0.4$}&\multicolumn{2}{c}{$d=0.47$} \\\hline
Method & \textit{c}  & \textit{SR} & \textit{CEQ}&  \textit{SR} & \textit{CEQ} & \textit{SR} & \textit{CEQ} & \textit{SR} & \textit{CEQ}\\\hline
Min regret  & 0.95  & 0 & 0 & 0  & 0 & 0
            &  0 & 0 & 0 \\
Soft 0.5 &   & 0 & 0&  0 & 0
  &  0&  0  & 0 & 0 \\
Borda  &   & \fbox{ 3 } &  \fbox{ 3 }  &  \fbox{ 3 }  & \fbox{ 3 } & \fbox{ 3 }  &  \fbox{ 3 }  & 2 & 2  \\
Copeland  &   & 2 &  2   &  1 & 1 &  1  & 1   & \fbox{ 3 } & \fbox{ 3 } \\\hline
Min regret    & 0.75  & \fbox{ 2 } &  0 & 1 &  0
        & 0 &  0  & 0  & 0\\
Soft 0.5 &   & 0 & 0 & 1   & 0  & 1
        &  0 &  2 &  0 \\
Borda  &  & 0 & \fbox{ 3 } & \fbox{ 2 }  & \fbox{ 3 }  & 1  & \fbox{ 3 }  & \fbox{ 3 } & \fbox{ 3 }  \\
Copeland  & & 1  & \fbox{ 3 }  & \fbox{ 2 } & \fbox{ 3 } & \fbox{ 2 }   & 2 & 0 & 1  \\\hline
Min regret   &0.50  & 0 & 0  & 1 & 0
       & 0 & 0 & 0 & 0\\
Soft 0.5 &   & 1 & 0 & 0  & 0  & 1 & 0 & \fbox{ 3 } & 1 \\
Borda  &   & \fbox{ 3 } &  \fbox{ 3 }   & \fbox{ 3 } & \fbox{ 3 } & \fbox{ 2 } & \fbox{ 3 } & 1 & \fbox{ 1 }\\
Copeland  &  & 2  & \fbox{ 3 }  & 1 & 2 & 1 & 2 & 0  & \fbox{ 1 }\\\hline
Min regret    &0.25  & 1 & 0   & \fbox{ 1 } & 0 & 0  & 0   & 0 & 0\\
Soft 0.5 &   & \fbox{ 2 } & 0 &  \fbox{ 1 }  & 0 & 1 & 0 & \fbox{ 3 } & \fbox{ 2 }   \\
Borda  &    & 1 & \fbox{ 3 }  & \fbox{ 1 }   & 2 & \fbox{ 2 } & \fbox{ 3 }  & 0 & 0\\
Copeland  &   &  0 & 1  & \fbox{ 1 }   & \fbox{ 3 } & 1  & \fbox{ 3 }  & 0  & 1 \\\hline

\multicolumn{10}{c}{} \\

\multicolumn{10}{l}{Panel B: Selected constituents of S\&P 100}       \\ \hline
\multicolumn{2}{c|}{} &
   \multicolumn{8}{c}{Kendall tau distance} \\

 \multicolumn{2}{c|}{} &
   \multicolumn{2}{c|}{$d=0.2$}&\multicolumn{2}{c|}{$d=0.3$}&\multicolumn{2}{c|}{$d=0.4$}&\multicolumn{2}{c}{$d=0.47$} \\\hline
Method & \textit{c}  & \textit{SR} & \textit{CEQ}&  \textit{SR} & \textit{CEQ} & \textit{SR} & \textit{CEQ} & \textit{SR} & \textit{CEQ}\\\hline
Min regret  & 0.95  & 0 & 0 & 0  & 0 & 0
            &  0 & 0 & 0 \\
Soft 0.5 &   & 0 & 0&  0 & 0
  &  0&  0  & 0 & 0 \\
Borda  &   & \fbox{ 3 } &  \fbox{ 3 }  &  \fbox{ 3 }  & \fbox{ 3 } & \fbox{ 3 }  &  \fbox{ 3 }  & 1 & 1  \\
Copeland  &   & 2 &  2   &  2 & 1 &  1  & 0   & \fbox{ 2 } & \fbox{ 2 } \\\hline
Min regret    & 0.75  & \fbox{ 2 } &  0 & \fbox{ 2 } &  0
        & 0 &  0  & 0  & 0\\
Soft 0.5 &   & 1 & 0 & 0   & 0  & 1
        &  0 &  1 &  0 \\
Borda  &  & 0 & \fbox{ 3 } & \fbox{ 2 }  & \fbox{ 3 }  & 1  & \fbox{ 3 }  & 0 & 0  \\
Copeland  & & 1  & \fbox{ 3 }  & \fbox{ 2 } & \fbox{ 3 } & \fbox{ 2 }   & 2 & \fbox{ 2 } & \fbox{ 3 }  \\\hline
Min regret   &0.50  & \fbox{ 3 } & 0  & \fbox{ 3 } & 0
       & \fbox{ 2 } & 0 & 0 & 0\\
Soft 0.5 &   & 0 & 0 & 0  & 0  & 1 & 0 & \fbox{ 2 } & 1 \\
Borda  &   & 0 &  \fbox{ 3 }   & 1 & \fbox{ 3 } & 1 & \fbox{ 3 } & 1 & 1\\
Copeland  &  & 0  & \fbox{ 3 }  & 1 & \fbox{ 3 } & \fbox{ 2 } & 2 & 0  & \fbox{ 2 }\\\hline
Min regret    &0.25  & \fbox{ 2 } & 0   & \fbox{ 2 } & 0 & \fbox{ 1 }  & 0   & 0 & 0\\
Soft 0.5 &   & 1 & 0 &  0  & 0 & \fbox{ 1 } & 0 & \fbox{ 3 } & \fbox{ 2 }   \\
Borda  &    & 0 & \fbox{ 3 }  & 1   & \fbox{ 3 } & \fbox{ 1 } & \fbox{ 3 }  & 0 & 0\\
Copeland  &   &  0 & 2  & 1   & \fbox{ 3 } & 0  & 2  & 1  & \fbox{ 2 } \\\hline

\end{tabular}

\caption{Number of wins for selected methods of aggregating qualitative views per different pairs of values  of the  Kendall tau distance $d$ and the  confidence parameter $c$. The numbers are aggregated across three different levels for the number of views $K\in \{5, 10, 20\} $. Results are counted as wins if they deviate from the optimum within a parameter setting by less than 1 percent.  \textit{SR} (\textit{CEQ}) denotes the Sharpe Ratio (Certainty-Equivalent return). Solid  boxes give the highest number of total wins within a parameter setting. Panel A (B) shows results for 38 (76)  stocks in the EURO STOXX 50 (S\&P 100). Sample period: January 1999 to December 2021. Source: Refinitiv Datastream.}
\label{tab:summary}
\end{table}

First, considering the results for different confidence levels $c$, the two methods based on robust optimization are reasonably competitive with the two social choice methods for higher confidence levels, i.e.\ $c\leq 0.5$, but are inferior for low confidence levels, e.g.\ not achieving a single win for $c=0.95$ in either sample.
This can be  explained by the fact that these methods follow a ``first estimate, then  aggregate'' philosophy, i.e., they first translate views into posterior  return estimators and then consider these estimators as scenarios in the context of the robust optimization model. 
For low confidence levels (high values of $c$), the information contained in the views has a minor impact on the posterior estimator of the vector of  expected returns, and the latter is not significantly different from the prior estimator. This  leads to very similar scenarios for the robust optimization. Consequently, the  optimal robust portfolio does not differ much from the optimal portfolio obtained by using only the prior estimator of expected returns and not considering any qualitative views at all. 
The social choice methods, on the other hand, aggregate the views into one ordering, still taking all information into account.
Only when the posterior return estimator is calculated does the dampening effect of low confidence come into play. Therefore, and for both performance measures, the two social choice methods clearly dominate the horse races for higher values of $c$.

For high confidence levels (i.e., low values of $c$), the result is not so clear-cut, at least when looking at the Sharpe Ratios. Evaluating all scenarios with $c\leq 0.5$, Min regret shows the highest number of wins for \textit{SR} in the S\&P 100 sample. But even in this setting, the social choice methods remain the predominant methods in the EURO STOXX 50 sample and in general when \textit{CEQ} is used as the performance measure.

\medskip
Second, regarding the influence of different degrees of information validity, expressed by different distances $d$ of the views from the correct order, there is almost no effect on the ranking of the different methods. For both samples and for both performance measures, the social choice methods achieve a higher number of wins than group (I) methods almost without exception. 
The respective overall winner, which in most cases is the Borda rule,  achieves first places regardless of the parameter $d$. Only for a high level of confidence in the views ($c=0.25$) do the robust optimization methods beat the social choice methods in some settings; however, the degree of information quality turns out to be irrelevant for this result.

\medskip
Finally, in addition to the influence of the parameters $c$ and $d$ on the rankings of the different methods of aggregating multiple qualitative views in MVO, we are interested in how the number of considered views $K$ (i.e., total orders ) affects the outcome of our horse races. 
To answer this question, we refer to the  results in Table~\ref{tab:allK}, where the actual values of \textit{SR} and \textit{CEQ} are given for various values of $K=5,10,20$ for each $d$ (averaged over all $c$) and for each $c$ (averaged over all $d$).
We find that changing the number of views considered has almost no influence on the overall winner of our horse races.
Although the levels of our performance measures increase with the number of views considered, as can well be expected, the rankings of the respective methods in the different settings remain largely unaffected. 
Aggregating the views before estimating the posterior distribution of portfolio returns turns out to be  superior to estimating the individual distributions  and then aggregating them, mostly regardless of the number of views considered. 
Within the winning group of social methods, Borda tends to perform slightly better than Copeland for $K=5$, while for larger $K$, e.g.\ $K=20$, these two social choice approaches are more or less on par.



To emphasize our main experimental result, we present the total number of  wins    for our selected methods for all 48 parameter settings in Table~\ref{tab:short}.
It shows that the social choice methods dominate the robust optimization methods in our experiments for both panels and for both performance metrics. 
We therefore recommend that if an investor considers to include several qualitative views into the BL portfolio optimization framework (s)he should apply one of our proposed methods from social choice theory, which commonly "aggregates first and then estimates" rather than the other way around.
The approach suggested by Borda should be preferred over the Copeland rule since it is the winner in three out of our four subtables.


\begin{landscape}

\begin{table}
\begin{tiny}
    
	\centering

\begin{tabular}{l|cc |cc|cc|cc||cc |cc|cc|cc}


\multicolumn{17}{l}{Panel A: Selected constituents of EURO STOXX 50}       \\ \hline
 \textbf{K = 5} &
   \multicolumn{2}{c|}{$d=0.2$}&\multicolumn{2}{c|}{$d=0.3$}&\multicolumn{2}{c|}{$d=0.4$}&\multicolumn{2}{c||}{$d=0.47$} & \multicolumn{2}{c|}{$c=0.95$}&\multicolumn{2}{c|}{$c=0.75$}&\multicolumn{2}{c|}{$c=0.50$}&\multicolumn{2}{c}{$c=0.25$} \\\hline
Method  & \textit{SR} & \textit{CEQ}&  \textit{SR} & \textit{CEQ} & \textit{SR} & \textit{CEQ} & \textit{SR} & \textit{CEQ} & \textit{SR} & \textit{CEQ}&  \textit{SR} & \textit{CEQ} & \textit{SR} & \textit{CEQ} & \textit{SR} & \textit{CEQ}\\\hline
Min regret  & 1.43 &0.11 & 1.40  & 0.10  & 1.23 & 0.08      & 0.44  & 0.02 

& 1.01 &0.05 & 1.17  & 0.08  & 1.15 & 0.09      & 1.16  & 0.09\\

Soft 0.5  & 1.23 & 0.10 & 1.15 & 0.09  & 1.06 & 0.08      & 0.53  & 0.03 

& 0.49 &0.03 & 1.14  & 0.08  & 1.17 & 0.10      & \fbox{1.18}  & 0.10\\

Borda  & \fbox{1.45} & \fbox{0.12} & \fbox{1.42}  & \fbox{0.12}  & \fbox{1.34} & \fbox{0.10}      & \fbox{0.57}  & \fbox{0.04}  

& \fbox{1.12} & \fbox{0.06} & \fbox{1.24}  & \fbox{0.10}  & \fbox{1.24} & \fbox{0.11}      & \dbox{1.17} & \fbox{0.11}\\

Copeland  & 1.42 & \dbox{0.12} & \dbox{1.41}  & 0.11  & 1.32 & 0.10      & 0.55  & 0.03  

& 1.10 & 0.06 & \dbox{1.23}  & 0.10  & 1.21 & 0.11      & 1.16  & \dbox{0.10}\\

\hline

\multicolumn{17}{c}{} \\
\hline

\textbf{K = 10} &
   \multicolumn{2}{c|}{$d=0.2$}&\multicolumn{2}{c|}{$d=0.3$}&\multicolumn{2}{c|}{$d=0.4$}&\multicolumn{2}{c||}{$d=0.47$} & \multicolumn{2}{c|}{$c=0.95$}&\multicolumn{2}{c|}{$c=0.75$}&\multicolumn{2}{c|}{$c=0.50$}&\multicolumn{2}{c}{$c=0.25$} \\\hline
Method  & \textit{SR} & \textit{CEQ}&  \textit{SR} & \textit{CEQ} & \textit{SR} & \textit{CEQ} & \textit{SR} & \textit{CEQ} & \textit{SR} & \textit{CEQ}&  \textit{SR} & \textit{CEQ} & \textit{SR} & \textit{CEQ} & \textit{SR} & \textit{CEQ}\\\hline
Min regret  & 1.44 &0.10 & 1.40  & 0.10  & 1.26 & 0.08      & 0.58  & 0.03 

& 1.05 &0.06 & 1.22  & 0.08  & 1.22 & 0.09      & 1.19  & 0.09\\

Soft 0.5  & 1.32 & 0.11 & 1.28 & 0.10  & 1.17 & 0.09      & 0.78  & 0.05 

& 0.57 &0.03 & 1.32  & 0.09  & 1.34 & 0.11      & \fbox{1.33}  & 0.11\\

Borda  & \dbox{1.51} & \fbox{0.12} & \fbox{1.52}  & \fbox{0.12}  & \fbox{1.46} & \dbox{0.11}      & 0.84  & 0.06  

& \fbox{1.27} & \fbox{0.07} & 1.37  & \fbox{0.11}  & \fbox{1.38} & \dbox{0.12}      & 1.30  & 0.12\\

Copeland  & \fbox{1.52} & \dbox{0.12} & 1.49  & \dbox{0.12}  & \dbox{1.46} & \fbox{0.11}      & \fbox{0.85}  & \fbox{0.06}  

& \dbox{1.27} & \dbox{0.07} & \fbox{1.39}  & \dbox{0.11}  & 1.34 & \fbox{0.12}      & 1.31  & \fbox{0.12}\\

\hline

\multicolumn{17}{c}{} \\
\hline

\textbf{K = 20} &
   \multicolumn{2}{c|}{$d=0.2$}&\multicolumn{2}{c|}{$d=0.3$}&\multicolumn{2}{c|}{$d=0.4$}&\multicolumn{2}{c||}{$d=0.47$} & \multicolumn{2}{c|}{$c=0.95$}&\multicolumn{2}{c|}{$c=0.75$}&\multicolumn{2}{c|}{$c=0.50$}&\multicolumn{2}{c}{$c=0.25$} \\\hline
Method  & \textit{SR} & \textit{CEQ}&  \textit{SR} & \textit{CEQ} & \textit{SR} & \textit{CEQ} & \textit{SR} & \textit{CEQ} & \textit{SR} & \textit{CEQ}&  \textit{SR} & \textit{CEQ} & \textit{SR} & \textit{CEQ} & \textit{SR} & \textit{CEQ}\\\hline
Min regret  & 1.46 &0.10 & 1.43  & 0.10  & 1.23 & 0.08      & 0.65  & 0.03 

& 1.04 &0.06 & 1.28  & 0.08  & 1.24 & 0.09      & 1.22  & 0.09\\

Soft 0.5  & 1.37 & 0.11 & 1.41 & 0.11  & 1.32 & 0.10      & 1.04  & 0.07 

& 0.83 &0.05 & \fbox{1.43}  & 0.11  & \fbox{1.42} & 0.12      & \fbox{1.47}  & 0.12\\

Borda  & \dbox{1.50} & \fbox{0.12} & \dbox{1.50}  & \fbox{0.12}  & \fbox{1.48} & \dbox{0.12}      & 1.09  & 0.08  

& \fbox{1.36} & \fbox{0.08} & \dbox{1.43}  & \fbox{0.12}  & 1.38 & 0.12      & 1.39  & \dbox{0.13}\\

Copeland  & \fbox{1.50} & \dbox{0.12} & \fbox{1.50}  & \dbox{0.12}  & \dbox{1.48} & \fbox{0.12}      & \fbox{1.10}  & \fbox{0.08}  

& \dbox{1.36} & \dbox{0.08} & \dbox{1.43}  & \dbox{0.12}  & 1.40 & \fbox{0.13}      & 1.39  & \fbox{0.13}\\
\hline

\multicolumn{17}{c}{} \\

\multicolumn{17}{l}{Panel B: Selected constituents of S\&P 100}       \\ \hline

\textbf{K = 5} &
   \multicolumn{2}{c|}{$d=0.2$}&\multicolumn{2}{c|}{$d=0.3$}&\multicolumn{2}{c|}{$d=0.4$}&\multicolumn{2}{c||}{$d=0.47$} & \multicolumn{2}{c|}{$c=0.95$}&\multicolumn{2}{c|}{$c=0.75$}&\multicolumn{2}{c|}{$c=0.50$}&\multicolumn{2}{c}{$c=0.25$} \\\hline
Method  & \textit{SR} & \textit{CEQ}&  \textit{SR} & \textit{CEQ} & \textit{SR} & \textit{CEQ} & \textit{SR} & \textit{CEQ} & \textit{SR} & \textit{CEQ}&  \textit{SR} & \textit{CEQ} & \textit{SR} & \textit{CEQ} & \textit{SR} & \textit{CEQ}\\\hline
Min regret  & \fbox{1.64} &0.13 & 1.61  & 0.12  & 1.54 & 0.11      & 0.58  & 0.03 

& 1.36 &0.07 & 1.39  & 0.10  & 1.33 & 0.11      & 1.28  & 0.11\\

Soft 0.5  & 1.34 & 0.12 & 1.30 & 0.11  & 1.27 & 0.10      & 0.78  & 0.06 

& 0.57 &0.03 & 1.46  & 0.11  & \dbox{1.35} & 0.13      & \dbox{1.31}  & 0.13\\

Borda  & 1.58 & \fbox{0.14} & \fbox{1.63}  & \fbox{0.14}  & \fbox{1.63} & \fbox{0.13}  & 0.81 & 0.06 
& \fbox{1.53} & \fbox{0.08} & 1.47  & \dbox{0.12}  & 1.34 & \fbox{0.13} & \fbox{1.32}  & \dbox{0.13}\\

Copeland  & 1.44 & 0.14 & \dbox{1.62}  & \dbox{0.13}  & \dbox{1.63} & 0.13  & \fbox{0.85}  & \fbox{0.06}  
& 1.50 & 0.07 & \fbox{1.49}  & \fbox{0.12}  & \fbox{1.35} & \dbox{0.13}  & \dbox{1.32}  & \fbox{0.13}\\
\hline

\multicolumn{17}{c}{} \\
\hline

\textbf{K = 10} &
   \multicolumn{2}{c|}{$d=0.2$}&\multicolumn{2}{c|}{$d=0.3$}&\multicolumn{2}{c|}{$d=0.4$}&\multicolumn{2}{c||}{$d=0.47$} & \multicolumn{2}{c|}{$c=0.95$}&\multicolumn{2}{c|}{$c=0.75$}&\multicolumn{2}{c|}{$c=0.50$}&\multicolumn{2}{c}{$c=0.25$} \\\hline
Method  & \textit{SR} & \textit{CEQ}&  \textit{SR} & \textit{CEQ} & \textit{SR} & \textit{CEQ} & \textit{SR} & \textit{CEQ} & \textit{SR} & \textit{CEQ}&  \textit{SR} & \textit{CEQ} & \textit{SR} & \textit{CEQ} & \textit{SR} & \textit{CEQ}\\\hline
Min regret  & \fbox{1.65} &0.12 & \dbox{1.64}  & 0.12  & \fbox{1.59} & 0.11      & 0.69  & 0.03 

& 1.44 &0.07 & 1.45  & 0.10  & 1.39 & 0.11      & 1.30  & 0.11\\

Soft 0.5  & 1.45 & 0.13 & 1.35 & 0.12  & 1.27 & 0.11      & 1.05  & 0.08 

& 0.75 &0.04 & 1.52  & 0.12  & \dbox{1.45} & 0.14      & \fbox{1.39}  & 0.14\\

Borda  & 1.64 & \fbox{0.14} & \fbox{1.66}  & \fbox{0.14}  & 1.57 & \fbox{0.14}  & 1.20 & 0.09 

& \dbox{1.70} & \fbox{0.09} & 1.59  & \dbox{0.13}  & \fbox{1.46} & 0.15 & 1.31  & \dbox{0.15}\\

Copeland  & 1.63 & \dbox{0.14} & \fbox{1.66}  & \dbox{0.14}  & \dbox{1.58} & \dbox{0.14}  & \fbox{1.23}  & \fbox{0.10} & \fbox{1.71} & \dbox{0.09} & \fbox{1.62}  & \fbox{0.13}  & \dbox{1.46} & \fbox{0.15}  & 1.32  & \fbox{0.15}\\
\hline

\multicolumn{17}{c}{} \\
\hline

\textbf{K = 20} &
   \multicolumn{2}{c|}{$d=0.2$}&\multicolumn{2}{c|}{$d=0.3$}&\multicolumn{2}{c|}{$d=0.4$}&\multicolumn{2}{c||}{$d=0.47$} & \multicolumn{2}{c|}{$c=0.95$}&\multicolumn{2}{c|}{$c=0.75$}&\multicolumn{2}{c|}{$c=0.50$}&\multicolumn{2}{c}{$c=0.25$} \\\hline
Method  & \textit{SR} & \textit{CEQ}&  \textit{SR} & \textit{CEQ} & \textit{SR} & \textit{CEQ} & \textit{SR} & \textit{CEQ} & \textit{SR} & \textit{CEQ}&  \textit{SR} & \textit{CEQ} & \textit{SR} & \textit{CEQ} & \textit{SR} & \textit{CEQ}\\\hline
Min regret  & \fbox{1.66} &0.13 & \dbox{1.64}  & 0.12  & 1.56 & 0.11      & 0.82  & 0.03 

& 1.45 &0.07 & 1.47  & 0.10  & 1.42 & 0.11      & 1.33  & 0.11\\

Soft 0.5  & 1.51 & 0.14 & 1.47 & 0.13  & 1.47 & 0.13      & 1.31  & 0.11 

& 1.08 &0.06 & \fbox{1.67}  & 0.14  & \dbox{1.52} & 0.15      & \fbox{1.48}  & 0.15\\

Borda  & \dbox{1.66} & \fbox{0.15} & \fbox{1.65}  & \fbox{0.15}  & \fbox{1.62} & \fbox{0.14}  & 1.50 & 0.12 

& \fbox{1.84} & \fbox{0.09} & 1.63  & \dbox{0.15}  & \fbox{1.53} & \dbox{0.16} & 1.44  & \dbox{0.16}\\

Copeland & \dbox{1.65} & \dbox{0.15} & \dbox{1.64}  & \dbox{0.14}  & \dbox{1.62} & \dbox{0.14}  & \fbox{1.52}  & \fbox{0.12}  
& \dbox{1.82} & \dbox{0.09} & \dbox{1.65}  & \fbox{0.15}  & \dbox{1.52} & \fbox{0.16}  & 1.44  & \fbox{0.16}\\

\hline

\end{tabular}

\caption{Average values of Sharpe Ratio (\textit{SR}) and Certainty Equivalent return (\textit{CEQ}) obtained from portfolio simulation runs computed by two robust optimization-based methods (Min regret; Soft 0.5) and two social choice-based methods (Borda; Copeland). 
$K\in \{5, 10, 20\}$ indicates the number of views.
The average is taken over the  values of \textit{SR} (\textit{CEQ}) obtained for each fixed value of the Kendall-Tau distance $d$ or the level of confidence in the views $c$ while varying the other parameter in the whole range of values, respectively: $d\in \{0.2, 0.3, 0.4, 0.47\} $, $c\in \{ 0.95, 0.75, 0.50, 0.25\}$. For each of the 48 combinations of parameters $c, d$ and $K$, solid boxes give the method with the highest value for \textit{SR} (\textit{CEQ}), while dashed boxes mark those methods deviating from the optimum by less than 1 percent.
Panel A (B) shows results for 38 (76)  stocks in the EURO STOXX 50 (S\&P 100). Sample period: January 1999 to December 2021. Source: Refinitiv Datastream.}

\label{tab:allK}
 \end{tiny}
 \end{table}
\end{landscape}

\begin{table}[]
	\footnotesize 
	\centering

\begin{tabular}{l|c|c}


\multicolumn{3}{l}{Panel A: Selected constituents of EURO STOXX 50}       \\ \hline

Method & Sharpe Ratio \textit{SR} & Certainty-Equivalent return \textit{CEQ}\\ \hline
Min regret & 6 & 0 \\
Soft $0.5$   & 16 & 3 \\
Borda & \fbox{ 30 } & \fbox{ 41 } \\
Copeland & \dbox{ 18 } & \dbox{ 32 } \\ \hline

\multicolumn{3}{c}{} \\

\multicolumn{3}{l}{Panel B: Selected constituents of S\&P 100}       \\ \hline
Method & Sharpe Ratio \textit{SR} & Certainty-Equivalent return \textit{CEQ}\\ \hline
Min regret & 17 & 0 \\
Soft $0.5$   & 11 & 3 \\
Borda & \dbox{ 18 } & \fbox{ 38 } \\
Copeland & \fbox{ 19 } & \dbox{ 35 } \\ \hline

	\end{tabular}
\caption{Number of wins for selected methods of aggregating qualitative views. The numbers are aggregated across three different levels for the number of views $K\in  \{5, 10, 20\}$. The input views  are total  orders of assets  with a  Kendall tau distance $d$ to the correct order of returns: $d\in  \{0.2, 0.3, 0.4, 0.47\}$.   $c\in \{ 0.95, 0.75, 0.50, 0.25\}$ is the   parameter representing the  confidence in the views. Thus, $48$ portfolio  simulation runs are performed for each method and the maximum achievable number of wins per method  is $48$.  Results are counted as wins if they deviate from the optimum by less than 1 percent. Solid (dashed) boxes give the highest (second highest) number of total wins within a parameter setting. Panel A (B) shows results for 38 (76)  stocks in the EURO STOXX 50 (S\&P 100). Sample period: January 1999 to December 2021. Source: Refinitiv Datastream.}
\label{tab:short}
\end{table}



 \section{Conclusions}
 \label{sec:conclusion}
 
While it is recognized in practice that relative rankings of future asset returns are easier to generate and generally more reliable than estimates of absolute returns, research on portfolio theory concerning this topic is rather modest. 
Moreover,  portfolio allocation decisions are increasingly made by teams requiring compromises among several different views. This effect has not yet been taken into account in the portfolio optimization literature.

In this paper we propose two general approaches that allow  the integration of multiple and potentially divergent views on the ranking of expected future stock returns in mean-variance (MV) portfolio optimization using the parametric Black-Litterman-framework (BL).
While methods from group (I) first generate an estimator of expected returns for each individual view and then process all these estimators in a robust MV portfolio optimization framework, the methods based on social choice theory (group (II)) first aggregate all orderings into one single consensus view which subsequently enters MV optimization. 
We are not aware of any previous literature that has combined social choice theory with MV portfolio optimization in the BL framework. 

Applying long-term in-sample analyses to stocks included in one of two broad stock indices and differentiating the views in terms of (i) the level of relevant information they contain, (ii) the level of confidence in the correctness of each generated view, and (iii) the number of views an investor considers in his/her portfolio decision, we find that portfolios built upon our proposed social choice methods (group (II)) clearly outperform portfolios based on robust optimization methods. 
We therefore recommend that if multiple views are to be considered in the portfolio allocation process, these views should be aggregated into a single view in a first step. Subsequently, this view can be used to derive a quantitative estimator for the posterior distribution of expected stock returns by applying the method proposed in \cite{CeHaMePfe2021JBF}, which allows a qualitative view to be integrated into the BL model. 
Eventually, MVO can be applied to this extended BL estimator of expected returns.

\medskip
Since we are (to the best of our knowledge) the first to combine methods from social choice theory with portfolio optimization, a number of questions arises for future research.
As a generalization of our ordinal framework, we might consider views represented by  partial orders. 
The extended BL model from \cite{CeHaMePfe2021JBF} could process partial orders as input, which could then be used as inputs to the robust optimization approaches.
However, aggregation of partial orders is typically not studied in the context of social choice methods. 
A further extension could be the combination of qualitative and quantitative information.
Clearly, this would be straightforward for robust portfolio optimization methods, after ``translating'' information of both types into return estimators.
Exploiting the potential of social choice methods for such a setting seems to be far less obvious.

\medskip
As a consequence of our work for practical decision making on asset allocations we would recommend practitioners to take a fresh view at the process of finding consensus between differing opinions (e.g., within an investment committee).
It is well-known that personal relations, hierarchies and anticipation of expected behaviour have a notable influence on the outcome of a group decision.
We propose to use the tools of robust optimization and social choice theory to formalize and objectify the consensus building process for decisions involving ordered preferences in portfolio optimization.

\subsection*{Acknowledgements}
We thank Christian Klamler for valuable input on social choice theory and Stefan Lendl for helping with computational issues.
Ulrich Pferschy acknowledges support by the field of excellence ``COLIBRI'' of the University of Graz.

\subsection*{Statements and Declarations}

{\bf Competing interests}

\noindent
The authors have no relevant financial or non-financial interests to disclose.

\noindent
The authors have no competing interests to declare that are relevant to the content of this article.



\newpage

 \bibliography{lit-hafner}

 \end{document}